\documentclass[pra,twocolumn,superscriptaddress]{revtex4}
\usepackage{amsfonts}
\usepackage{amsmath}
\usepackage{amssymb}
\usepackage{color}
\usepackage{graphicx}
\usepackage{epsfig,subfigure,dsfont,amsthm,amsbsy,mathrsfs,amscd}
\usepackage{booktabs}
\newtheorem{theorem}{Theorem}

\begin{document}
\title{\large\bf Converting quantum coherence to genuine multipartite entanglement and nonlocality}

\author{Ya Xi}
\affiliation{Department of Mathematics, South China
University of Technology, GuangZhou 510640, China}

\author{Tinggui Zhang}
\email{tinggui333@163.com}
\affiliation{School of Mathematics and Statistics, Hainan Normal University, Haikou 571158, China}

\author{Zhu-Jun Zheng}
\affiliation{Department of Mathematics, South China University of Technology, GuangZhou 510640, China}

\author{Xianqing Li-Jost}

\affiliation{School of Mathematics and Statistics, Hainan Normal
University, Haikou 571158, China}
\affiliation{Max-Planck-Institute
for Mathematics in the Sciences, Leipzig 04103, Germany}
\author{Shao-Ming Fei}
\email{feishm@cnu.edu.cn}

\affiliation{Max-Planck-Institute for Mathematics in the Sciences,
Leipzig 04103, Germany} \affiliation{School of Mathematics and
Sciences, Capital Normal University, Beijing 100048, China}

\begin{abstract}
We study the relations between quantum coherence and quantum
nonlocality, genuine quantum entanglement and genuine quantum nonlocality.
We show that the coherence of a qubit state can be converted to the nonlocality of
two-qubit states via incoherent operations. The results are also generalized to
qudit case. Furthermore, rigorous relations
between the quantum coherence of a single-partite state and the genuine multipartite quantum entanglement, as well as
the genuine three-qubit quantum nonlocality are established.
\end{abstract}

\maketitle

\section{Introduction}

Quantum entanglement is a crucial resource for many quantum information processing tasks such as quantum teleportation, dense coding, and quantum key distribution; see \cite{mule1} for a review.
In particular, the genuinely multipartite entangled states \cite{guhnerev} offer significant advantages in quantum tasks comparing with bipartite entanglement. They are the basic ingredients in measurement-based quantum computation \cite{mule2}, and are beneficial in various quantum communication protocols \cite{mule4,hillery,srensen}.

Besides the quantum entanglement, the quantum nonlocality is also of great importance in both understanding the conceptual foundations of quantum theory and quantum information processing such as building quantum protocols to decrease communication complexity \cite{dcc,dcc1}
and providing secure quantum communication \cite{scc1,scc2}.

Comparing with quantum entanglement and quantum nonlocality which are defined for bipartite or multipartite systems, quantum coherence can be defined for single systems. Due to the superposition principle in quantum mechanics, it plays important roles in many researches such as quantum computation \cite{Hill16,Matera16,Giovannetti04}, quantum metrology \cite{EschMD11, MarvS16,Asboth05}.

It has been shown that coherence and quantum correlations can be converted to each other under certain scenarios. In \cite{StreSDB15}, the authors show that any degree of coherence in some reference basis can be converted to entanglement via incoherent operations. In \cite{zhuhj},
the authors further show that any entanglement of bipartite pure states is the minimum of a suitable coherence measure over product bases and conversely, any coherence measure
of pure states, with the extension to mixed states by the convex roof theory, is equal to the maximum entanglement generated by incoherent operations acting on the system and an incoherent ancilla. In \cite{MaYGV16}, the authors prove that the creation of quantum discord with multipartite incoherent operations is bounded by the consumption of quantum coherence in its subsystems.
In \cite{MSP2018,TCKJ2018} the authors show the conversion between quantum coherence and other convex resources including quantum fisher information and the ``magic" ones.

In this paper, we study the relations among quantum coherence, genuine quantum entanglement,
quantum nonlocality and genuine quantum nonlocality. We show that any nonzero coherence
can be converted to genuine multipartite entanglement, quantum nonlocality and genuine multipartite nonlocality under incoherent operations.

\section{Converting coherence to bipartite nonlocality}

The coherence of a quantum state depends on the reference basis. Throughout the paper, we fix the reference basis to be the computational basis. Let $H_d$ denote a $d$-dimensional Hilbert space.
The $l_1$-norm coherence $\mathcal{C}_{l_1}(\rho^s)$ of a source quantum state $\rho^s\in H_d$ is defined by \cite{t baumgratz},
\begin{equation}\label{cl1}
\mathcal{C}_{l_1}(\rho^s)=\sum_{i\neq j} |\rho_{ij}^s|,
\end{equation}
where $|\rho_{ij}^s|$ denotes the absolute value of $\rho_{ij}^s$.
The set of incoherent states is defined by
$\mathcal{I}\equiv\{\rho=\sum_{i=0}^{d-1}p_{i}|i\rangle\langle i| | \sum_{i=0}^{d-1}p_{i}=1\}.$
And a completely positive and trace-preserving map $\Lambda$ is called an incoherent operation if it can be written as
$\Lambda(\rho)=\sum_{j}K_{j}\rho K_{j}^{\dag},$ where every $K_{j}$ is incoherent in the sense that $K_{j}\mathcal{I}K_{j}^{\dag}\subseteq\mathcal{I}.$

Concerning the nonlocality, we first consider two-qubit state ${\rho}\equiv\rho^s\otimes |0\rangle\langle 0|\in H_2\otimes H_2$, where $\rho^s$ is the source qubit state and $|0\rangle\langle 0|$ is the initial state of an auxiliary qubit system. Let $\Lambda$ be an incoherent operation on $H_2\otimes H_2$.
We have the following result:

\begin{theorem}
The state $\Lambda(\rho)$ is Bell-nonlocal if and only if $\rho^s$ has non-vanishing coherence.
\end{theorem}

[{\sf Proof}]. Under computational basis, $\rho^s=\sum_{i,j=0}^{1}\rho_{ij}^s|i\rangle\langle j|$. Taking $\Lambda$ to be the two-qubit CNOT gate, $\Lambda|i\rangle\otimes|j\rangle=|i\rangle\otimes|mod(i+j,2)\rangle$,
then $\Lambda({\rho})=\sum_{jk}\rho^s_{jk}|jj\rangle\langle kk|$.
Denote $T$ the $3\times 3$ real matrix with entries given by $t_{nm}=Tr(\Lambda({\rho})\sigma_{n}\otimes\sigma_{m})$,
where $\{\sigma_{n}\}_{n=1}^{3}$ are the standard Pauli matrices.
Set $U\equiv T^{t}T$ with $t$ standing for the transposition of the matrices.
Let $\mu_1$ and ${\mu}_2$ be the two larger eigenvalues of $U$ and
$M(\Lambda({\rho}))\equiv \mu_1+{\mu}_2$.

A state is non-local if it violates any Bell inequalities.
For the two qubit state $\Lambda({\rho})$, the CHSH inequality says that
$|\langle\mathbb{B}_{CHSH}\rangle_{\Lambda({\rho})}|\leq 2$,
where $\mathbb{B}_{CHSH}=A_1\otimes B_1+A_1\otimes B_2+A_2\otimes
B_1-A_2\otimes B_2$, $\langle\mathbb{B}_{CHSH}\rangle_{\Lambda({\rho})}=Tr(\Lambda({\rho})\,\mathbb{B}_{CHSH})$, $A_i=\vec{a}_i\cdot\vec{\sigma}$,
$B_j=\vec{b}_j\cdot\vec{\sigma}$, $i,j=1,2$, $\vec{\sigma}=(\sigma_1,\sigma_2,\sigma_3)$, $\vec{a}_i$ and $\vec{b}_j$ are unit real three dimensional vectors.
The state $\Lambda({\rho})$ violates the CHSH inequality if $M(\Lambda({\rho}))>1$ \cite{3H1995}.

It is direct to verify that $M(\Lambda({\rho}))=1+4|{\rho}_{01}^s|^2$.
Therefore, the quantum state $\Lambda({\rho})$ is non-local if $|{\rho}_{01}^s|\neq 0$, namely,
if ${\rho}^s$ has non-zero coherence. $\Box$

\begin{figure}[ptb]
\includegraphics[width=0.4\textwidth]{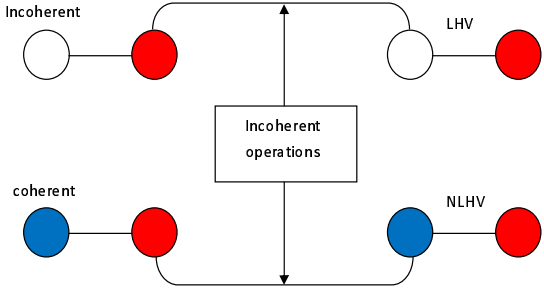}
\caption{One to one mapping between quantum coherence and quantum nonlocality under incoherent transformations.
For initial incoherent states, under incoherent operations the resulting states admit local hidden variable (LHV) models.
While for coherent states, the resulting states admit no local hidden variable (NLHV) models}
\label{Fig2}%
\end{figure}

Theorem 1 says that as long as the source state $\rho^s$ has non-zero coherence,
there exist incoherent operations such that the coherence in the state $\rho^s$ can be converted to the Bell nonlocality of the state $\Lambda({\rho})$, see Fig. \ref{Fig2}.
As nonlocality necessarily implies entanglement, our conclusion is stronger than the one in \cite{StreSDB15}, in which any degree of coherence can be converted to entanglement via incoherent operations.

In order to study the relations between the coherence and quantum
nonlocality for high dimensional bipartite states $\rho\in
H_d\otimes H_d$, we need the following fact. Let $|e_\alpha\rangle$,
$\alpha=0,1,2,\cdots,d-1$, be a basis of $H_d$. Set $P=P_A\otimes
P_B$, where $P_A=(|e_\alpha\rangle,|e_\beta\rangle)^t$ for some
$\alpha\neq \beta$, and $P_B=(|e_\gamma\rangle,|e_\lambda\rangle)^t$
for some $\gamma\neq \lambda$. Then
$$
\widetilde{\rho}=\frac{P\rho P^{\dag}}{Tr[P\rho P^{\dag}]}
$$
is a ``two-qubit" state. Since
$P_A^{\dag}\sigma_iP_A$ and $P_B^{\dag}\sigma_iP_B$ have the same
eigenvalues as $\sigma_i$, the state $\rho$ is non-local if
$|\langle{\mathbb{B}}_{CHSH}\rangle_{\tilde{\rho}}|>\frac{2}{Tr[P\rho P^{\dag}]}$. That is, if
$|\langle\widetilde{\mathbb{B}}_{CHSH}\rangle_{\rho}|>2$, then
$\rho$ is non-local, where
$\widetilde{\mathbb{B}}_{CHSH}=P^{\dag}\mathbb{B}_{CHSH}P$ is the
CHSH operator induced from $\mathbb{B}_{CHSH}$.

\begin{theorem} For any source state $\rho^{s}=\sum_{ij}\rho_{ij}^{s}|i\rangle\langle j|\in H_{d}$, if for some $i\neq j$, $|\rho_{ij}^{s}|>\frac{\sqrt{1-(\rho_{ii}^{s}+\rho_{jj}^{s})^{2}}}{2},$ then the state $\rho^s\otimes |0\rangle\langle 0|\in H_{d}\otimes H_d$ can be converted to be Bell-nonlocal via incoherent operations $\Lambda$ such that $\Lambda(\rho^{s}\otimes|0\rangle\langle 0|)$ violates a Bell-inequality.
\end{theorem}

[{\sf Proof}]. Note that $\max|\langle\widetilde{\mathbb{B}}_{CHSH}\rangle_{\rho}|=\max|Tr[\widetilde{\mathbb{B}}_{CHSH}\, \rho]|
=\max Tr[P\rho P^{\dag}]\,|Tr[\mathbb{B}_{CHSH}\widetilde{\rho}]|=2Tr[P\rho P^{\dag}]\sqrt{M(\widetilde{\rho})}$. Associated with the corresponding projector $P$, $Tr[P\Lambda(\rho) P^{\dag}]=\rho_{ii}^{s}+\rho_{jj}^{s}$ and $\sqrt{M(\widetilde{\Lambda(\rho)})}=\sqrt{(\rho_{ii}^{s}+\rho_{jj}^{s})^{2}+4|\rho_{ij}^{s}|^2}/({\rho_{ii}^{s}+\rho_{jj}^{s}})$ for some $i\neq j$. Then $2(\rho_{ii}^{s}+\rho_{jj}^{s})\sqrt{M(\widetilde{\Lambda(\rho)})}>2$ if and only if $|\rho_{ij}^{s}| > \frac{\sqrt{1-(\rho_{ii}^{s}+\rho_{jj}^{s})^{2}}}{2} $.
Therefore, if $|\rho_{ij}^{s}| >\frac{\sqrt{1-(\rho_{ii}^{s}+\rho_{jj}^{s})^{2}}}{2}$, the state $\rho^s\otimes |0\rangle\langle 0|$ can be converted to a Bell-nonlocal state via incoherent
operations such that $\Lambda(\rho^{s}\otimes|0\rangle\langle 0|)$ violates a Bell-inequality. $\Box$

In particular, if $\rho^s$ is a rank two state, $\rho_{kk}^{s}>0$, $\rho_{ll}^{s}> 0$ and $\rho_{kk}^{s}+\rho_{ll}^{s}=1$ for some $k\neq l$, the $|\rho_{kl}^{s}|>0$ if $\rho^{s}$ is coherent.
Taking $P_{A}=P_{B}=(|e_k\rangle,|e_l\rangle)^t$, one gets $Tr[P\Lambda(\rho) P^{\dag}]\,\sqrt{M(\widetilde{\Lambda(\rho)})}=\sqrt{1+4|\rho_{kl}^{s}|^2} > 1$. By
the proof of Theorem 2, we have the following necessary and sufficient conclusion:

{\bf Corollary.}~
If the source state $\rho^{s}\in H_{d}$ has rank two, then the state $\rho^s\otimes |0\rangle\langle 0|$ can be converted to be Bell-nonlocal under incoherent
operations if and only if $\rho^s$ is coherent.

Based on the von Neumann relative entropy $S(\rho\parallel\sigma)$ of two quantum states $\rho$ and $\sigma$, in \cite{ZYLF2017} the authors studied a unified characterization of quantum correlations for entanglement $\mathcal{E}(\rho)=\min\limits_{\sigma\in \mathcal{S}}S(\rho\parallel \sigma)$,
discord $\mathcal{D}(\rho)=\min\limits_{\sigma\in \mathcal{CC}}S(\rho\parallel\sigma)$,
steerability $\mathcal{\widehat{S}}(\rho)=\min\limits_{\sigma\in \mathcal{U}}S(\rho\parallel \sigma)$,
nonlocality $\mathcal{N}(\rho)=\min\limits_{\sigma\in \mathcal L}S(\rho\parallel\sigma)$, as well as
the coherence $\mathcal{C}_r(\rho)=\min\limits_{\sigma\in\mathcal{I}}S(\rho\parallel \sigma)$,
where $\mathcal{S}$, $\mathcal{CC}$, $\mathcal{U}$ and $\mathcal{L}$ stand
for the sets of separable states, classically correlated states, unsteerable states,
the states admitting local hidden variable models, respectively. Since $\mathcal{I}\subset\mathcal{CC}\subset \mathcal{S}\subset\mathcal{U}\subset\mathcal{L}$,
one has the relation $\mathcal{C}_r(\rho)\geq\mathcal{D}(\rho) \geq \mathcal{E}(\rho) \geq
\mathcal{\widehat{S}}(\rho) \geq \mathcal{N}(\rho)$.
Let $Q$ denote one of the correlations $\mathcal{E}$, $\mathcal{D}$, $\mathcal{\widehat{S}}$ and $\mathcal{N}$. We have the following general conclusion.

\begin{theorem}
The quantum correlation of a bipartite state
$\Lambda(\rho^{s}\otimes|0\rangle\langle0|^{A})$ generated by
incoherent operation $\Lambda$ on an initial source qudit state
$\rho^{s}$ and an ancilla state $|0\rangle\langle0|^{A}$ of system
$A$ is upper bounded by the relative entropy coherence $\rho^{s}$,
\begin{equation}\label{thm2}
Q(\Lambda(\rho^{s}\otimes|0\rangle\langle0|^{A}))\leq
\mathcal{C}_{r}(\rho^{s}).
\end{equation}
\end{theorem}

[{\sf Proof}]. Let $\sigma^{s}$ be the closet incoherent state to
$\rho^{s}$, i.e.
$\mathcal{C}_{r}(\rho^{s})=S(\rho^{s}\|\sigma^{s}).$ And
$S(\rho^{s}\|\sigma^{s})=S(\rho^{s}\otimes|0\rangle\langle0|^{A}\|\sigma^{s}\otimes|0\rangle\langle0|^{A})$ \cite{StreSDB15}. Based on the monotonicity of $S$ under completely positive operations $\Lambda$, $S(\Lambda(\rho)\|\Lambda(\sigma))\leq S(\rho\|\sigma)$, we have
$S(\rho^{s}\otimes|0\rangle\langle0|^{A}\|\sigma^{s}\otimes|0\rangle\langle0|^{A})\geq S(\Lambda(\rho^{s}\otimes|0\rangle\langle0|^{A})\|\Lambda(\sigma^{s}\otimes|0\rangle\langle0|^{A}))$.
Since $\Lambda(\sigma^{s}\otimes|0\rangle\langle0|^{A})$ is an incoherent state, it is also classically correlated, separable, unsteerable and local. Therefore, we get
$\mathcal{C}_{r}(\rho^{s})=S(\rho^{s}\|\sigma^{s})=S(\rho^{s}\otimes|0\rangle\langle0|^{A}\|\sigma^{s}\otimes|0\rangle\langle0|^{A})\geq
S(\Lambda(\rho^{s}\otimes|0\rangle\langle0|^{A})\|\Lambda(\sigma^{s}\otimes|0\rangle\langle0|^{A}))
\geq\mathcal{C}_{r}(\Lambda(\rho^{s}\otimes|0\rangle\langle0|^{A})\geq
Q(\Lambda(\rho^{s}\otimes|0\rangle\langle0|^{A})).$   $\Box$

Theorem 3 provides a lower bound of $\mathcal{C}_{r}(\rho^{s})$, given by the distance-based quantum correlations $Q$ between the source state and ancilla state under incoherent operations.
In fact, due to the relation between the relative entropy of coherence and the $l_1$-norm coherence, $\mathcal{C}_{r}(\rho^{s})\leq \log_{2}(d)\mathcal{C}_{l_{1}}(\rho^{s})$ \cite{RPL2016}, (\ref{thm2}) gives rise to  $Q(\Lambda(\rho^{s}\otimes|0\rangle\langle0|^{A}))\leq\log_{2}(d)\mathcal{C}_{l_{1}}(\rho^{s})$,
which presents also an upper bound of quantum correlation measures by the $l_1$-norm coherence.

\section{Converting coherence to genuine tripartite entanglement and nonlocality}

\noindent {\sf Coherence to genuine tripartite entanglement}~~
Different from the ones in bipartite systems, states in tripartite systems can be not only entangled
or non-locally correlated, but also genuinely entangled or genuinely non-locally correlated.
Genuine multipartite entanglement is an important type of entanglement which offers
significant advantages in quantum tasks comparing with bipartite
entanglement \cite{mule1}. It is also the basic ingredient
in measurement-based quantum computation \cite{mule2}, and in various quantum communication protocols \cite{mule3}
including secret sharing \cite{mule44,mule5}.
A genuinely multipartite entangled mixed
state is defined to be one that cannot be written as a convex
combination of bi-separable pure states.
We consider general three-qudit case.
Let $\rho^{s}$ be the qudit source state, $|0\rangle\langle0|^{A}$ and $|0\rangle\langle0|^{B}$ the initial states of the auxiliary qudits $A$ and $B$, respectively.

\begin{theorem}
The state $\rho^{s}\otimes|0\rangle\langle0|^{A}\otimes|0\rangle\langle0|^{B}$ can be converted to a genuinely
tripartite entangled state under incoherent operations if and only if $\rho^{s}$ is coherent.
\end{theorem}

[{\sf Proof}].
If there exist incoherent operations such that $\rho^{s}\otimes|0\rangle\langle0|^{A}\otimes|0\rangle\langle0|^{B}$ is converted to a genuinely tripartite entangled one, by Theorem 3 $\rho^{s}$ must be coherent.
Conversely, consider the incoherent unitary operator
$\mathcal{U}=\Sigma_{i=0}^{d-1}\Sigma_{j=0}^{d-1}\Sigma_{k=0}^{d-1}|i\rangle\langle i|\otimes|mod(i+j,d)\rangle\langle j|\otimes|mod(i+k,d)\rangle\langle k|$.
We have $\mathcal{U}(\rho^{s}\otimes|0\rangle\langle0|^{A}\otimes|0\rangle\langle0|^{B})
=\Sigma_{ij}\rho_{ij}^{s}|i\rangle\langle j|\otimes|i\rangle\langle j|\otimes|i\rangle\langle j|.$
If $\mathcal{U}(\rho^{s}\otimes|0\rangle\langle0|^{A}\otimes|0\rangle\langle0|^{B})$ is a pure state, using the genuine mutipartite concurrence $C_{gme}(|\Psi\rangle)=\min\limits_{C\in\{C,\bar{C}\}}\sqrt{{2(1-Tr(Tr_{C}(|\Psi\rangle\langle \Psi|))^{2})}}$ \cite{GME2011}, where $\{C,\bar{C}\}$ denotes bipartite decompositions of $|\Psi\rangle$, we get $C_{gme}(\mathcal{U}(\rho^{s}\otimes|0\rangle\langle0|^{A}\otimes|0\rangle\langle0|^{B}))=2\sqrt{\Sigma_{k\neq l}\rho^{s}_{kk}\rho^{s}_{ll}}$.
This genuine mutipartite concurrence is great than zero if $\mathcal{C}_{l_1}(\rho^s)\neq0$, since in this case there must exist some $k$ and $l$ such that $|\rho_{kl}^s|\neq 0$, namely,
$\rho^{s}_{kk}\neq 0$ and $\rho^{s}_{ll}\neq 0$ due to the positivity of a density matrix.
Therefore, $\mathcal{U}(\rho^{s}\otimes|0\rangle\langle0|^{A}\otimes|0\rangle\langle0|^{B})$ is a genuine tripartite entangled state.
If the state $\mathcal{U}(\rho^{s}\otimes|0\rangle\langle0|^{A}\otimes|0\rangle\langle0|^{B})$ is a mixed one,
$\mathcal{U}(\rho^{s}\otimes|0\rangle\langle0|^{A}\otimes|0\rangle\langle0|^{B})=\mathcal{U}(\Sigma_{k}p_{k}|\psi_{k}\rangle\langle \psi_{k}|^{s}\otimes|0\rangle\langle0|^{A}\otimes|0\rangle\langle0|^{B})=\Sigma_{k}p_{k}\mathcal{U}(|\psi_{k}\rangle\langle \psi_{k}|^{s}\otimes|0\rangle\langle0|^{A}\otimes|0\rangle\langle0|^{B})$. If $C_{l_1}(\rho^s)\neq0$, given the convexity of coherence measures, there will exist some $\mathcal{U}(|\psi_{k}\rangle\langle \psi_{k}|^{s}\otimes|0\rangle\langle0|^{A}\otimes|0\rangle\langle0|^{B})$ which is coherent. By the above proof about pure states, we can get $C_{gme}(\mathcal{U}(\rho^{s}\otimes|0\rangle\langle0|^{A}\otimes|0\rangle\langle0|^{B})=\min\limits_{\{p_{k},|\psi_{k}\rangle\}}\Sigma_{k}p_{k}C_{gme}(\mathcal{U}(|\psi_{k}\rangle\langle \psi_{k}|^{s}\otimes|0\rangle\langle0|^{A}\otimes|0\rangle\langle0|^{B}))>0$. Then $\rho^{s}\otimes|0\rangle\langle0|^{A}\otimes|0\rangle\langle0|^{B}$ can be converted to a genuinely tripartite entangled state via incoherent operations if and only if $\rho^{s}$ is coherent. $\Box$

{\it Remark} 1) When $\mathcal{C}_{l_{1}}(\rho^{s})=1$, i.e., $|\rho_{01}^s|={1}/{2}$,
$\mathcal{U}(\rho^{s}\otimes|0\rangle\langle0|^{A}\otimes|0\rangle\langle0|^{B})$
is just the GHZ state, $(|000\rangle+|111\rangle)/{\sqrt{2}}$.

2) For arbitrary $\rho^{s}$ and incoherent operation $\mathcal{U}$, it is impossible that $\mathcal{U}(\rho^{s}\otimes|0\rangle\langle0|^{A}\otimes|0\rangle\langle0|^{B})$ is the state $|W\rangle=
(|001\rangle+|010\rangle+|100\rangle)/\sqrt{3}$. The reason is that $\mathcal{C}_{l_1}(\mathcal{U}(\rho^{s}\otimes|0\rangle\langle0|^{A}\otimes|0\rangle\langle0|^{B}))\leq \mathcal{C}_{l_1}(\rho^{s}\otimes|0\rangle\langle0|^{A}\otimes|0\rangle\langle0|^{B})=\mathcal{C}_{l_1}(\rho^{s})$, but $\mathcal{C}_{l_1}(|W\rangle\langle W|)=2>\mathcal{C}_{l_1}(\rho^{s}),$ where $\mathcal{C}_{l_1}(\rho^{s})=2|\rho_{01}^s|\leq2\sqrt{\rho_{00}^s\rho_{11}^s}\leq2\sqrt{(\frac{\rho_{00}^s+\rho_{11}^s}{2})^{2}}=1$ by the fact that $\rho^{s}$ is semipositive definite with trace one.

In particular, by the similar method with Theorem 4, it is direct to get that the state $\rho^{s}$ can be converted to a genuinely multipartite entangled state under incoherent operations if and only if $\rho^{s}$ is coherent. In order to illustrate this, we will give the following Example.

[{\it Example}]
Consider the n-qubit quantum state  $\Lambda(\rho^{s}\otimes|0\rangle\langle0|\otimes\cdots\otimes|0\rangle\langle0|)$, where $\Lambda=\Sigma_{i=0}^{1}\Sigma_{j=0}^{1}\cdots\Sigma_{k=0}^{1}|i\rangle\langle i|\otimes|mod(i+j,2)\rangle\langle j|\otimes\cdots\otimes|mod(i+k,2)\rangle\langle k|$,
then $\Lambda(\rho^{s}\otimes|0\rangle\langle0|\otimes\cdots|0\rangle\langle0|)
=\Sigma_{i,j}\rho_{ij}^{s}|i\rangle\langle j|\otimes|i\rangle\langle j|\otimes\cdots\otimes|i\rangle\langle j|.$
The genuine mutipartite concurrence is given by \cite{GME-X}, $C_{gme}(\Lambda(\rho^{s}\otimes|0\rangle\langle0|\otimes\cdots\otimes|0\rangle\langle0|))=2|\rho_{01}^{s}|$. Namely, $\rho^{s}$ can be converted to an n-qubit genuine entangled state if and only if $\rho^{s}$ is a coherent state.

\noindent {\sf Coherence to genuine tripartite nonlocality}~~
Quantum nonlocality can be revealed via violations of various Bell
inequalities.
It has been recognized that quantum nonlocality is not only a puzzling aspect of nature, but also an important resource for quantum information processing,
such as building quantum protocols to decrease communication complexity \cite{dcc,dcc1}
and providing secure quantum communication \cite{scc1,scc2}.
For tripartite case, there are so called genuine tripartite nonlocality and three-way nonlocal correlations.
If Alice, Bob and Charlie perform measurement $X$, $Y$ and $Z$ on the three subsystems, respectively, with
outcomes $x$, $y$ and $z$, and the probability correlations $P(xyz|XYZ)$ among the measurement outcomes
can be written as in the hybrid local-nonlocal form \cite{S1987},
$P(xyz|XYZ)=\sum_{\lambda}q_{\lambda} P_{\lambda}(xy|XY) P_{\lambda}(z|Z)
+\sum_{\mu}q_{\mu} P_{\mu}(xz|XZ) P_{\mu}(y|Y)
+\sum_{\nu}q_{\nu} P_{\nu} (yz|YZ)P_{\nu} (x|X)$,
where $0\leq q_{\lambda}$, $q_{\mu}, q_{\nu} \leq 1$ and $\sum_{\lambda}q_{\lambda}+\sum_{\mu}q_{\mu}+\sum_{\nu}q_{\nu}=1$,
then the correlations are called Svetlichny local. Otherwise we call them genuinely Svetlichny nonlocal.
Concerning the coherence and the genuine three qubit nonlocality, we have the following conclusion.

\begin{theorem}
If $\mathcal{C}_{l_{1}}(\rho^{s})>\sqrt{\frac{1}{2}}$, then
the state $\rho^{s}\otimes|0\rangle\langle0|^{A}\otimes|0\rangle\langle0|^{B}$
can be converted to be a genuinely three-qubit nonlocal state via incoherent operations.
\end{theorem}

[{\sf Proof}].
It has been shown that a three-qubit state $|\Psi\rangle$ admits bi-local hidden variable model if the mean value of the Svetlichny operator $\mathcal{S}$ is bounded \cite{S1987},
$|\langle\Psi|\mathcal{S}|\Psi\rangle|\leq 4$. While in \cite{LSJFL2017} the authors show that for any three-qubit quantum state $\rho$,
the maximal mean value of the Svetlichny operator $\mathcal{S}$  satisfies: $\max|\langle\mathcal{S}\rangle_{\rho}|\leq 4\lambda_{1}$,
where $\lambda_{1}$ is the maximum singular value of the matrix $m=(m_{j,ik})$,
with $m_{ijk}=Tr\left[\rho(\sigma_{i}\otimes\sigma_{j}\otimes\sigma_{k})\right]$, $i,j,k=1,2,3$.
It is easily calculated that
$\lambda_{1}(\mathcal{U}(\rho^{s}\otimes|0\rangle\langle0|^{A}\otimes|0\rangle\langle0|^{B}))=2\sqrt{2}|\rho_{01}^{s}|=\sqrt{2} \mathcal{C}_{l_{1}}(\rho^{s})$.
Therefore, if $\mathcal{C}_{l_{1}}(\rho^{s})>\sqrt{\frac{1}{2}}$,
$\mathcal{U}(\rho^{s}\otimes|0\rangle\langle0|^{A}\otimes|0\rangle\langle0|^{B})$ is a genuinely three-qubit nonlocal state. $\Box$

The violation of the Svetlichny inequality is only a sufficient condition of genuine three-qubit nonlocalilty.
In \cite{Nonlocal2013} other three-qubit genuine nonlocality, three-way nonlocal correlations, have been studied.
Let us consider the $\mathcal{T}$ inequality, $\langle\mathcal{T}\rangle\equiv\langle X_{0}Y_{0}\rangle+\langle X_{0}Z_{0}\rangle+\langle Y_{0}Z_{1}\rangle-\langle X_{1}Y_{1}Z_{0}\rangle +\langle X_{1}Y_{1}Z_{1}\rangle\leq3$, and the $\mathcal{NS}$ inequality, $\langle\mathcal{NS}\rangle\equiv\langle X_{0}Y_{1}\rangle+\langle X_{1}Z_{0}\rangle+\langle Y_{0}Z_{1}\rangle+\langle X_{0}Y_{0}Z_{0}\rangle -\langle X_{1}Y_{1}Z_{1}\rangle\leq3$
given in \cite{Nonlocal2013}, where $\langle X_{i}Y_{j}Z_{k}\rangle=\sum_{xyz}(-1)^{x+y+z}p(xyz|X_{i}Y_{j}Z_{k})$ and $X_{i}$, $Y_{j}$, $Z_{k}$, $i,j,k=0,1$, have binary outcomes $x,y,z.$
By numerical calculation, we can see the violation of these two inequalities, see Fig. 2.

\begin{figure}[!htbp]
\includegraphics[width=0.4\textwidth]{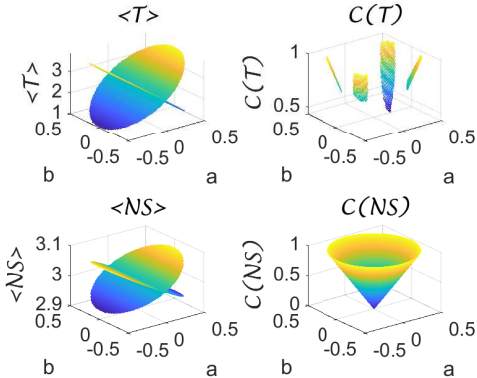}
\caption{The surfaces of $\langle\mathcal{T}\rangle$, $\langle\mathcal{NS}\rangle$, $\mathcal{C}(\mathcal{T})$ and $\mathcal{C}(\mathcal{NS})$ with respect to state $\rho^{s}$.
$\langle\mathcal{T}\rangle$ and $\langle\mathcal{NS}\rangle$ are the corresponding values of the state
$\mathcal{U}(\rho^{s}\otimes|0\rangle\langle0|^{A}\otimes|0\rangle\langle0|^{B})$, $\mathcal{C}(\mathcal{T})$ ($\mathcal{C}(\mathcal{NS})$) is the coherence of $\rho^{s}$ such that $\mathcal{U}(\rho^{s}\otimes|0\rangle\langle0|^{A}\otimes|0\rangle\langle0|^{B})$ violate the $\mathcal{T}$ ($\mathcal{NS}$) inequality, and $a+i b=\rho_{01}^{s}$.}\label{Fig3}
\end{figure}

In fact, for $\langle\mathcal{T}\rangle$,  taking $X_{0}=Y_{0}=\sigma_{3}$, $X_{1}=Y_{1}=\sigma_{1}$, $Z_{0}=\frac{1}{\sqrt{2}}(\sigma_{3}-\sigma_{1})$, $Z_{1}=\frac{1}{\sqrt{2}}(\sigma_{3}+\sigma_{1})$, we have $\langle\mathcal{T}\rangle$ of $\mathcal{U}(\rho^{s}\otimes|0\rangle\langle0|^{A}\otimes|0\rangle\langle0|^{B}):$ $\langle\mathcal{T}\rangle=1+\sqrt{2}+\sqrt{2}(\rho_{01}^{s}+\rho_{10}^{s})=1+\sqrt{2}+2\sqrt{2}a$. And taking $X_{0}=Y_{0}=\sigma_{3}$, $X_{1}=Y_{1}=\sigma_{1}$, $Z_{0}=\frac{1}{\sqrt{2}}(\sigma_{3}+\sigma_{1})$, $Z_{1}=\frac{1}{\sqrt{2}}(\sigma_{3}-\sigma_{1})$, we have $\langle\mathcal{T}\rangle=1+\sqrt{2}-\sqrt{2}(\rho_{01}^{s}+\rho_{10}^{s})=1+\sqrt{2}-2\sqrt{2}a.$ Therefore, from $\langle\mathcal{T}\rangle>3$, we get $|a|>\frac{\sqrt{2}-1}{2}$. Namely, if $|a|>\frac{\sqrt{2}-1}{2}$, the state $\mathcal{U}(\rho^{s}\otimes|0\rangle\langle0|^{A}\otimes|0\rangle\langle0|^{B})$ violates the $\mathcal{T}$ inequalities.
Similarly, we have that if $|a|>0,$ the state $\mathcal{U}(\rho^{s}\otimes|0\rangle\langle0|^{A}\otimes|0\rangle\langle0|^{B})$ violates the $\mathcal{NS}$ inequalities.

We summarize the corresponding lower bounds of $\mathcal{C}_{l_{1}}(\rho^{s})$ that under incoherent operation $\mathcal{U}$
the state $\rho^{s}$ can be converted to $\mathcal{U}(\rho^{s}\otimes|0\rangle\langle0|^{A}\otimes|0\rangle\langle0|^{B})$ which
is genuinely tripartite entangled (GME), the genuinely Svetlichny ($\mathcal{S}$), $\mathcal{T}$ and $\mathcal{NS}$ nonlocal in Table 1.

\begin{table}[!hbp]
  \centering
  \begin{tabular}{|c|c|c|c|c|}
  \hline
&$\mathcal{S}$& $\mathcal{T}$& $\mathcal{NS}$ & GME \\
\hline
$\mathcal{C}_{l_{1}}(\rho^{s}) $& ${1}/{\sqrt{2}}$& $\sqrt{2}-1$&0& 0\\
    \hline
\end{tabular}
\caption{The lower bounds of $\mathcal{C}_{l_{1}}(\rho^{s})$ exhibiting genuine three qubit entanglement and violating the Svetlichny, $\mathcal{T}$ and $\mathcal{NS}$ inequalities.}\label{1}
  \end{table}

\section{ Conclusions and discussions}
We have established rigorous relations
between the conversion of  quantum coherence and bipartite nonlocality, tripartite genuine entanglement and genuine nonlocality.
It has been proven that any nonzero coherence of a source qubit state $\rho^s$ can be converted to the Bell-nonlocality under incoherent transformations.
Moreover, any arbitrary dimensional tripartite state $\rho^{s}\otimes|0\rangle\langle0|^{A}\otimes|0\rangle\langle0|^{B}$ can be converted to a genuinely tripartite entangled state under incoherent operations as long as $\rho^{s}$ is coherent.
And when $\mathcal{C}_{l_{1}}(\rho^{s})\in (\sqrt{\frac{1}{2}},1]$, i.e., $|\rho_{01}^{s}|\in (\frac{1}{2\sqrt{2}},\frac{1}{2}]$, the state
$\rho^{s}\otimes|0\rangle\langle0|^{A}\otimes|0\rangle\langle0|^{B}$ can be converted to a genuinely three-qubit nonlocal state.
While to exhibit the NS three way nonlocality, nonzero coherence of $\rho^s$ suffices.
Besides the one to one correspondence between coherence and entanglement, our results show further the tight relations between the coherence
and quantum nonlocality, genuine tripartite entanglement and genuine tripartite nonlocality.

\bigskip
\bigskip
\noindent{\bf Acknowledgments}\, \,  We are grateful to the referees for constructive suggestions. This project is supported by the China Scholarship Council, the National Natural Science Foundation of China (Grants No.11571119, No.11861031, 11675113 and 11501153) and Beijing Municipal Commission of Education (KZ201810028042).

\end{document}